\documentclass{article}
\usepackage{graphicx}
\usepackage{amsthm}
\usepackage{amsmath}
\usepackage{amsfonts}
\usepackage{amssymb}

\newtheorem{thm}{Theorem}

\newtheorem{proposition}[thm]{Proposition}

\theoremstyle{definition}
\newtheorem{definition}{Definition}
\theoremstyle{remark}
\newtheorem{remark}{Remark}

\begin{document}
\baselineskip=19pt

\title{Vacuum Einstein metrics with\\bidimensional Killing leaves.\footnote{
Published in Diff. Geom. Appl. 17(2002)15-35.
\newline Research supported in part by the
Italian Ministero dell' Universit\`{a} e della Ricerca Scientifica e
Tecnologica.} \\\textit{II-Global aspects}.}
\author{G. Sparano $^{\text{\#,\S}}$, \ G. Vilasi $^{\text{\#,\S\S}}$, A. M.
Vinogradov $^{\text{\#,\S}}$\\{\small \ }$^{\text{\#}}${\small \
Istituto Nazionale di Fisica Nucleare, Gruppo Collegato di
Salerno, Italy.}\\{\small \ }$^{\text{\S}}${\small \ Dipartimento
di Matematica e Informatica, Universit\`{a} di Salerno, Italy.
}\\{\small \ }$^{\text{\S\S}}${\small \ Dipartimento di Scienze
Fisiche \textit{E.R.Caianiello}, Universit\`{a} di Salerno, Italy.
}\\{\small \ (e-mail: sparano@unisa.it, vilasi@sa.infn.it,
vinogradov@ponza.dia.unisa.it)}} \maketitle \vspace{1.0cm}

\begin{abstract}

A formalism ($\zeta $-{\it complex analysis}), allowing one to construct
global Einstein metrics by matching together local ones  described in the
papers [Phys. Lett. \textbf{B} 513(2001)142-146;  Diff. Geom. Appl.
\textbf{16}(2002)95-120], is developed. With  this formalism the
singularities of the obtained metrics are  described naturally as well.\
{\it Subj. Class.: }Differential Geometry, General Relativity.    {\it
Keywords: }Einstein metrics, Killing vectors.    {\it MS classification:
}53C25
\newpage
\end{abstract}

\newpage

\section{Introduction}

In the previous paper \cite{1SVV00} we described local form of vacuum
Einstein metrics that admit a Killing algebra  $\mathcal{G}$, such that
\begin{description}
\item [ I]the distribution $\mathcal{D}$, generated by the vector fields
belonging to $\mathcal{G}$, is bidimensional
\item[ II] the distribution $\mathcal{D}^{\bot}$, orthogonal to $\mathcal{D}$,
is completely integrable and transversal to $\mathcal{D}$.
\end{description}

In this paper we answer the natural question: how to match together these
local metrics, in order to get \textit{global nonextendable} ones. An
important common peculiarity of the considered local metrics is that they
all  are fibered over $\zeta$\textit{-complex curves} (see next section).
This fact  allows one to reduce the problem to a much simple one in
zeta-complex  analysis. Moreover, singularities of Einstein metrics, which
are inevitable  according to Hawking`s theorem, may be described with this
technique in a  simple transparent manner. The same technique offers also
various  possibilities in manipulating with already known Einstein metrics
to get new  ones. For instance, just by ''extracting the square roots''
from the  Schwarzschild metric one discovers two ''parallel universes''
generated by two  ''parallel stars''.    Geometrical properties of
solutions described in the paper will be discussed  with more details
separately. Here, we will limit ourselves to a few examples  (section
\ref{esempi}) illustrating some aspects of our approach which  generalizes
naturally to several situations as, for instance,
\textit{cosmological Einstein metrics }satisfying assumptions I and II (work
in progress).   In this paper we continue to use terminological and
notational  conventions, adopted in \cite{1SVV00}, without a special
mentioning.
\section{Global solutions\label{global}}

From the local analytic description of Ricci-flat metrics, given in
\cite{1SVV00} (section \textit{5} and \textit{6)}, it is not immediately
evident whether they are pair-wise different or not. Here we will give them
a  coordinate-free description, so that it becomes clear what variety of
different geometries, in fact, is obtained.    We will see that with any of
the found solutions a pair, consisting of a  $\zeta$\textit{-complex curve}
$\mathcal{W}$ and a $\zeta$\textit{-harmonic}  function $u$ on it, is
associated. If two solutions are equivalent, then the  corresponding pairs,
say $\left(  \mathcal{W},u\right)  $ and $\left(
\mathcal{W}^{\prime},u^{\prime}\right)  $, are related by an invertible
$\mathbb{\zeta}$-holomorphic map $\Phi:\left(  \mathcal{W},u\right)
\longrightarrow\left(  \mathcal{W}^{\prime},u^{\prime}\right)  $ such that
$\Phi^{\ast}\left(  u^{\prime}\right)  =u$. Roughly speaking, the
\textit{\ moduli space }of the obtained geometries is surjectively mapped on
the \textit{moduli space }of the pairs $\left(  \mathcal{W},u\right)  $.
Further parameters, distinguishing the metrics we are analyzing, are given
below. Before that, however, it is worth to underline the following common
peculiarities of these metrics.
\begin{proposition}
Orthogonal leaves are totally geodesic and possess a non trivial Killing
field. Geodesic flows, corresponding to metrics, admitting \textit
{3}-dimensional  Killing algebras, are non-commutatively integrable.
\end{proposition}
\label{gfi}

\begin{proof}
The first assertion follows from the fact that the metrics have, in the
adapted coordinates, a block diagonal form whose upper block does not
depend  on the last two coordinates. The existence of a non trivial Killing
field is  obvious from the description of model solution given in next
section. For what  concerns geodesic flows, they are integrated explicitly
for model solution in  next section, and the general result follows from
the fact that any solution  is a pullback of a model one.
\end{proof}

Solutions of the Einstein equations found in \cite{1SVV00} (section
\textit{5}) manifest an interesting common feature. Namely, each of them is
determined completely by a choice of
\begin{itemize}
\item [$1)$]a solution of the wave, or the Laplace equation, depending on the
sign of $\det\mathbf{F}$;
\end{itemize}

and either by
\begin{itemize}
\item [$2^{\prime})$]a choice of the constant $A$ and one of the branches (see
fig.1 in section \textit{\ref{gps}}) of the tortoise equation
\begin{equation}
\beta+A\ln\left|  \beta-A\right|  =u,\label{sp1}
\end{equation}
if $h_{22}\neq0$, or by
\item[$2^{\prime\prime})$] a choice of a solution of one of the two equations
\begin{gather}
\left[  \mu\left(  \partial_{y}^{2}-\partial_{x}^{2}\right)  +\mu_{y}
\partial_{y}-\mu_{x}\partial_{x}\right]  w=0\qquad\square\mu=0,\label{h02}\\
\left[  \mu\left(  \partial_{y}^{2}+\partial_{x}^{2}\right)  +\mu_{y}
\partial_{y}+\mu_{x}\partial_{x}\right]  w=0\qquad\triangle\mu=0,\label{l02}
\end{gather}
according to the sign of $\det\mathbf{F}$, in the case $h_{22}=0.$
\end{itemize}

Solutions we are analyzing in the paper have a natural \textit{fibered
structure} with the Killing leaves as fibers. The wave and Laplace
equations,  mentioned above in $1)$, are in fact defined on the
bidimensional manifold  $\mathcal{W}$ which parameterizes the Killing
leaves. These leaves themselves  are bidimensional Riemannian manifolds
and, as such, are geodesically complete.    For this reason the problem of
the extension of local solutions, found in  \cite{1SVV00} (section
\textit{5})\textit{,} is reduced to that of the  extension of the base
manifold $\mathcal{W}$. Such an extension should carry a  geometrical
structure that gives an intrinsic sense to the notion of the wave  or the
Laplace equation and to equations (\ref{h02}) and (\ref{l02}) on it. A
brief description of how this can be done is the following.
\subsection{$\mathbb{\zeta}-$complex structures}

Recall that there exist three different isomorphism classes of
bidimensional commutative unitary algebras. They are
\[
\mathbb{C}=\mathbb{R}\left[  x\right]  /\left(  x^{2}+1\right)  ,\qquad
\mathbb{R}_{\left(  2\right)  }=\mathbb{R}\left[  x\right]  /\left(
x^{2}\right)  ,\qquad\mathbb{R}\oplus\mathbb{R}=\mathbb{R}\left[  x\right]
/\left(  x^{2}-1\right)  .
\]

Elements of this algebra can be represented in the form $a+\mathbb{\zeta}
b,$ $a,b\in\mathbb{R}$, with $\mathbb{\zeta}^{2}=-1,0,$ or $1$,
respectively. For  a terminological convenience we will call them
$\mathbb{\zeta}$
\textit{-complex numbers}. Of course, $\mathbb{\zeta}$-complex numbers for
$\mathbb{\zeta}^{2}=-1$ are just ordinary complex numbers. Furthermore, we
will use the unifying notation $\mathbb{R}_{\mathbb{\zeta}}^{2}$ for the
algebra of $\mathbb{\zeta}$-complex numbers. For instance $\mathbb{C}=$
$\mathbb{R}_{\mathbb{\zeta}}^{2}$ for $\mathbb{\zeta}^{2}=-1$.
\begin{remark}
In the literature one can find various alternative names for elements of
these algebras. For instance, \textit{dual} numbers  for
$\mathbb{R}_{\left( 2\right) }$, \textit{double} or
\textit{Zarissky} numbers for $\mathbb{R}\oplus \mathbb{R}$, etc.
\end{remark}

In full parallel with ordinary complex numbers, it is possible to develop a
$\mathbb{\zeta}$\textit{-complex analysis }by defining $\mathbb{\zeta}
$\textit{-holomorphic} functions as
$\mathbb{R}_{\mathbb{\zeta}}^{2}$-valued  differentiable functions of the
variable $z=x+\mathbb{\zeta} y$. Just as in  the case of ordinary complex
numbers, the function $f\left(  z\right)
=u\left(  x,y\right)  $+$\mathbb{\zeta} v\left(  x,y\right)  $ is
$\mathbb{\zeta}$-holomorphic\textit{\ iff} the
$\mathbb{\zeta}$-Cauchy-Riemann conditions hold:
\begin{equation}
u_{x}=v_{y},\qquad u_{y}=\mathbb{\zeta}^{2}v_{x}\label{pcr}
\end{equation}
\textit{\ }The compatibility conditions of the above system requires that both
$u$ and $v$ satisfy the $\mathbb{\zeta}$\textit{-Laplace equation}, that is

\[
-\mathbb{\zeta}^{2}u_{xx}+u_{yy}=0,\qquad-\mathbb{\zeta}^{2}v_{xx}+v_{yy}=0.
\]
Of course, the $\mathbb{\zeta}$\textit{-}Laplace\textit{\ }equation reduces
for $\mathbb{\zeta}^{2}=-1$ to the ordinary Laplace equation, while for
$\mathbb{\zeta}^{2}=1$ to the wave equation. The operator $-\mathbb{\zeta}
^{2}\partial_{x}^{2}+\partial_{y}^{2}$ will be called the $\mathbb{\zeta}
$\textit{-Laplace operator}.
\begin{definition}
(i) A $\zeta$-complex structure on $\cal W$ is an endomorphism
$J:\cal{D}\left({\cal W}\right)\rightarrow\cal{D}\left({\cal W}\right  )$
of the  $C^\infty\left({\cal W}\right)$ module $\cal{D}\left({\cal
W}\right)$ of all  vector fields on $\cal{W}$, with
$J^{2}=\zeta^{2}\mathbf{I}$, $J\neq  0,\mathbf{I}$,  and vanishing
Nijenhuis  torsion, i.e.,  $\left[J,J\right]^{FN}=0$, where $\left[ ,
\right]^{FN}$ stands for
the Fr\"{o}licher-Nijenhuis bracket. (ii) A bidimensional manifold  $\cal W$
supplied  with a {\it$\zeta$-complex structure} is called a
{\it$\zeta$-complex curve}.
\end{definition}

Obviously, for $\zeta^{2}=-1$ a $\mathbb{\zeta}$-complex curve\textit{\ }
is just an ordinary \textit{1}-dimensional complex manifold (curve).    By
using the endomorphism $J$ the $\mathbb{\zeta}$-Laplace equation can be
written intrinsically as
\[
d\left(  J^{\ast}du\right)  =0,
\]
where $J^{\ast}:\Lambda^{1}\left(  \mathcal{W}\right)  $ $\longrightarrow
\Lambda^{1}\left(  \mathcal{W}\right)  $ is the \textit{adjoint to }
$J$\textit{\ endomorphism} of the $C^{\infty}\left(  \mathcal{W}\right)  $
module of $1$-forms on $\mathcal{W}$.    Given a bidimensional smooth
manifold $\mathcal{W}$, an atlas $\left\{
\left(  U_{i},\Phi_{i}\right)  \right\}  $ on $\mathcal{W}$ is called
$\mathbb{\ \zeta}$\textit{-complex } \textit{iff}
\begin{itemize}
\item [ i)]$\Phi_{i}:U_{i}\longrightarrow\mathcal{W},\quad U_{i}$ is open in
$\mathbb{R}_{\mathbb{\zeta}}^{2}$,
\item[ ii)] the transition functions $\Phi_{j}^{-1}\circ\Phi_{i}\ $are
$\mathbb{\zeta}$-holomorphic.
\end{itemize}

Two $\mathbb{\zeta}$-complex atlases on $\mathcal{W}$ are said to be
\textit{\ equivalent} if their union is again a $\mathbb{\zeta}$-complex atlas.

A class of $\mathbb{\zeta}$-complex atlases on $\mathcal{W}$ supplies,
obviously, $\mathcal{W}$ with a $\mathbb{\zeta}$-complex
structure\textit{.}  Conversely, given a $\mathbb{\zeta}$-complex structure
on $\mathcal{W}$ there  exists a $\mathbb{\zeta}$-complex atlas on
$\mathcal{W}$ inducing this  structure. Charts of such an atlas will be
called $\mathbb{\zeta}   $\textit{-complex coordinates }on the
corresponding\textit{\ }$\mathbb{\zeta}  $-complex curve. In
$\mathbb{\zeta}$-complex coordinates the endomorphism $J $  and its adjoint
$J^{\ast}$ are described by the relations
\begin{align*}
J\left(  \partial_{x}\right)   & =\partial_{y,\,\,\,\,\,\,}J\left(
\partial_{y}\right)  =\mathbb{\zeta}^{2}\partial_{x}\\
J^{\ast}\left(  dx\right)   & =\mathbb{\zeta}^{2}dy_{,\,\,\,\,\,\,}J^{\ast
}\left(  dy\right)  =dx.
\end{align*}

If $\mathbb{\zeta}^{2}\neq0$, the functions $u$ and $v$ in the Eq.
(\ref{pcr}) are said to be \textit{conjugate}.    Alternatively, a
$\mathbb{\zeta}$-complex curve can be regarded as a  bidimensional smooth
manifold supplied with a specific atlas whose transition  functions
\[
(x,y)\longmapsto\left(  \mathbb{\xi}(x,y),\mathbb{\eta}(x,y)\right)
\]
are subjected to $\mathbb{\zeta}$-Cauchy-Riemann relations (\ref{pcr}).
\begin{remark}
It is not difficult to see that for $\mathbb{\zeta}^{2}=1$ a
$\mathbb{\zeta}$
-complex structure on a bidimensional manifold is completely determined by
its characteristic distribution, \textit{i.e.}, by two
\textit{1}-dimensional distributions composed of characteristic vectors of
the corresponding  $\zeta$-Laplace equation, and conversely.
\end{remark}
.

As it is easy to see, the $\mathbb{\zeta}$-Cauchy-Riemann relations
(\ref{pcr}) imply that
\[
\partial_{\mathbb{\eta}}^{2}-\mathbb{\zeta}^{2}\partial_{\mathbb{\xi}}
^{2}=\frac{1}{\mathbb{\xi}_{x}^{2}-\mathbb{\zeta}^{2}\mathbb{\xi}_{y}^{2}
}\left(  \partial_{y}^{2}-\mathbb{\ \zeta}^{2}\partial_{x}^{2}\right)  ,
\]
and also
\[
\mu\left(  \partial_{\mathbb{\eta}}^{2}-\mathbb{\zeta}^{2}\partial
_{\mathbb{\xi}}^{2}\right)  +\mu_{\mathbb{\eta}}\partial_{\mathbb{\eta}
}-\mathbb{\zeta}^{2}\mu_{\mathbb{\xi}}\partial_{\mathbb{\xi}}=\frac
{1}{\mathbb{\xi}_{x}^{2}-\mathbb{\zeta}^{2}\mathbb{\xi}_{y}^{2}}\left[
\mu\left(  \partial_{y}^{2}-\mathbb{\ \zeta}^{2}\partial_{x}^{2}\right)
+\mu_{y}\partial_{y}-\mathbb{\zeta}^{2}\mu_{x}\partial_{x}\right]  .
\]

This shows that equation (\ref{h02}) (respectively, (\ref{l02})) is
well-defined on a $\mathbb{\zeta}$-complex curve with $\mathbb{\zeta}^{2}=1
$  (respectively, $\mathbb{\zeta}^{2}=-1$). The manifestly intrinsic
expression  for these equations is
\[
d\left(  \mu J^{\ast}dw\right)  =0.
\]
We will refer to it as the $\mu$-\textit{deformed }$\mathbb{\zeta}
$\textit{-Laplace equation.}    A solution of the $\mathbb{\zeta}$-Laplace
equation on $\mathcal{W}$ will be  called
$\mathbb{\zeta}$\textit{-harmonic}.\textit{\ }We can see that in the  case
\textit{\ }$\mathbb{\zeta}^{2}\neq0$ the notion of \textit{conjugate
}$\zeta$\textit{-harmonic function} is well defined on a $\mathbb{\zeta}
$-complex curve. In addition, notice that the metric field $d\mathbb{\xi}
^{2}-\mathbb{\zeta}^{2}d\mathbb{\eta}^{2}$, $\mathbb{\eta}$ being
$\mathbb{\zeta}$-conjugate with $\xi$, is canonically associated with a
$\mathbb{\zeta}$-harmonic function $\xi$ on $\mathcal{W}$.    A map
$\Phi:\mathcal{W}_{1}\longrightarrow\mathcal{W}_{2}$ connecting two
$\mathbb{\zeta}$-complex curves will be called $\mathbb{\zeta}$
\textit{-holomorphic }if $\varphi\circ\Phi$ is locally $\mathbb{\zeta}
$-holomorphic for any local $\mathbb{\zeta}$-holomorphic function $\varphi$
on $\mathcal{W}_{2}$. Obviously, if $\Phi$ is $\mathbb{\zeta}$-holomorphic
and  $u$ is a $\mathbb{\zeta}$ -harmonic function on $\mathcal{W}_{2}$,
then  $\Phi^{\ast}\left(  u\right)  $ is $\mathbb{\zeta}$ -harmonic on
$\mathcal{W}_{1}$.
\begin{itemize}
\item  The \textit{standard }$\zeta$\textit{-complex curve} is $\mathbb{R}
_{\mathbb{\zeta}}^{2}=\left\{  \left(  x+\mathbb{\zeta} y\right)  \right\}  $,
and the \textit{standard }$\zeta$\textit{-harmonic function} on it is given
by $x$, whose conjugated is $y$.
\end{itemize}

The pair $\left(  \mathbb{R}_{\mathbb{\zeta}}^{2},x\right)  $ is
\textit{universal } in the sense that for a given $\mathbb{\zeta}$-harmonic
function $u$ on a $\mathbb{\zeta}$-complex curve $\mathcal{W}$ there exists
a $\mathbb{\zeta}$-holomorphic map $\Phi:\mathcal{W}\longrightarrow
\mathbb{R}_{\mathbb{\zeta}}^{2}$ defined uniquely by the relations $\Phi
^{\ast}\left(  x\right)  =u$ and $\Phi^{\ast}\left(  y\right)  =v$, $v$ being
conjugated with $u$.
\subsection{Global properties of solutions\label{gps}}

The above discussion shows that any global solution, that can be obtained
by matching together local solutions found in \cite{1SVV00} (section
\textit{5}),  is a solution whose base manifold is a
$\mathbb{\zeta}$-complex curve  $\mathcal{W}$ and which corresponds to a
$\mathbb{\zeta}$-harmonic function  $u$ on $\mathcal{W}$.    A solution of
Einstein equations corresponding to $\mathcal{W}\subseteq$
$\mathbb{R}_{\mathbb{\zeta}}^{2}$, $u\equiv x$ will be called  a\
\textit{model}. Notice that there exist various model solutions due to
various options in the choice of parameters appearing in $2^{\prime})$ and
$2^{\prime\prime})$ at the beginning of this section. An important role
played  by the model solutions is revealed by the following assertion:
\begin{proposition}
\label{pb}Any solution of the Einstein equation which can be
constructed by matching together local solutions of \cite{1SVV00} (section
5) is the pullback of a model solution \textit{via} a
$\mathbb{\zeta}$-holomorphic map from a $\mathbb{\zeta}$-complex  curve to
$\mathbb{R}_{\mathbb{\zeta}}^{2}.$
\end{proposition}

\begin{proof}
It follows directly from the fact that any $\mathcal{\zeta}$-harmonic
function on a $\mathcal{\zeta}$-complex curve $\mathcal{W}$ is the pullback
of $x$ by a  suitable $\zeta$-holomorphic map $\mathcal{W}\longrightarrow
R_{\mathcal{\zeta  }}^{2}.$
\end{proof}

Now we can resume in a systematic way the results obtained in the first
part \cite{1SVV00}.    We distinguish between the two following
qualitatively different cases:
\begin{itemize}
\item [\textit{I}]metrics admitting a normal \textit{3-}dimensional Killing
algebra with bidimensional leaves;
\item[\textit{II}] metrics admitting a normal bidimensional Killing algebra
that does not \textit{extend} to a larger algebra having the same leaves
and whose distribution orthogonal to the leaves is integrable.
\end{itemize}

It is worth mentioning that the distribution orthogonal to the Killing
leaves is automatically integrable in \textit{Case I} (proposition
\textit{16 }of  \cite{1SVV00}, section \textit{7}). Also in \textit{Case
II} the  bidimensionality of the Killing leaves is guaranteed by
proposition \textit{2}  of \cite{1SVV00}.    Any Ricci-flat manifold
$\left(  M,g\right)  $, we are analyzing, is fibered  over a
$\mathbb{\zeta}$-complex curve $\mathcal{W}$
\[
\pi:M\longrightarrow\mathcal{W}
\]
whose fibers are the Killing leaves and as such are bidimensional Riemann
manifolds of constant Gauss curvature.    Below, we shall call $\pi$ the
\textit{Killing fibering }and assume that its  fibers are \textit{connected
and geodesically complete}. Therefore, maximal  (\textit{i.e.,}
\textit{non-extendible}) Ricci-flat manifolds, of the class we  are
analyzing in the paper, are those corresponding to maximal (\textit{i.e.,}
\textit{non-extendible}) pairs $\left(  \mathcal{W},u\right)  $, where
$\mathcal{W}$ is a $\mathbb{\zeta}$-complex curve and $u$ is
$\mathbb{\zeta}  $-harmonic function on $\mathcal{W}$.
\subsubsection{\textit{Case I}}

Here the Killing algebra $\mathcal{G}$ is isomorphic to one of the
following: $so\left(  3\right)  $, $so\left(  2,1\right)  $,
$\mathcal{K}\mathit{il}
\left(  dx^{2}\pm dy^{2}\right)  $, $\mathcal{A}_{3}$ (see section \textit{7}
in \cite{1SVV00}), and the Killing fibering splits in a canonical way into
the Cartesian product.    This product structure can be interpreted as a
flat connection in $\pi$,  determined uniquely by the requirements that its
parallel sections are  orthogonal to the Killing leaves and the parallel
transports of fibers are  their conform equivalences (with respect to the
induced metrics). In that  sense one can say that the Killing fibering is
supplied canonically with a  conformally flat connection. This is a
geometrically intrinsic way to describe  the Killing fibering.    A
discrete isometry group acting freely on a fiber of the Killing fibering
can  be extended fiber-wise to the whole of $M$ due to the canonical
product  structure mentioned above. Conversely, a locally isometric
covering  $\psi:\widetilde{S}\longrightarrow S$ of a fiber $S$ allows to
construct a  locally isometric covering $\widetilde{M}\longrightarrow M$
which copies  $\psi$ fiber by fiber. So any homogeneous bidimensional
Riemannian manifold  can be realized as typical fiber of a Killing
fibering.    Denote by $\left(  \Sigma,g_{\Sigma}\right)  $ a homogeneous
bidimensional  Riemannian manifold, whose Gauss curvature $K\left(
g_{\Sigma}\right)  $, if  different from zero, is normalized to $\pm1$.
Denote by $\left(
\mathcal{W},u\right)  $ the pair constituted by a $\mathbb{\zeta}$-complex
curve $\mathcal{W}$ and a $\mathbb{\zeta}$-harmonic function $u$ on
$\mathcal{W}$. Denote also by $\pi_{1}$ (respectively, $\pi_{2}$ ) the
natural  projection of $M=\mathcal{W} \Sigma$ on $\mathcal{W}$
(respectively, on  $\Sigma$). Then, the above data determine the following
Ricci- flat manifold  $\left(  M,g\right)  $ with
\begin{equation}
M=\mathcal{W} \Sigma,\qquad g=\pi_{1}^{\ast}\left(  g_{\left\{  u\right\}
}\right)  +\pi_{1}^{\ast}\left(  \beta^{2}\right)  \pi_{2}^{\ast  }\left(
g_{\Sigma}\right) \label{form}
\end{equation}
where $\beta=\beta\left(  u\right)  $ is implicitly determined by $u$ via
the equation
\begin{equation}
\beta+A\ln\left|  \beta-A\right|  =u\label{bal}
\end{equation}
and
\begin{equation}
g_{\left\{  u\right\}  }=\epsilon\frac{\beta-A}{\beta}\left(  du^{2}
-\mathbb{\ \zeta}^{2}dv^{2}\right) \label{met}
\end{equation}
$A$ being an arbitrary constant and $\epsilon=\pm1$ .   Only in the case
$A=0$ the equation (\ref{bal}) determines the function  $\beta\left(
u\right)  $ uniquely: $\beta\equiv u$ and $g$ is flat.    If $A\neq0$ the
graph of the left hand side of Eq.(\ref{bal}) is as follows
\[
\begin{array}
[c]{cc} {\includegraphics[ natheight=2.000300in, natwidth=2.332400in,
height=2.0003in, width=2.3324in]   {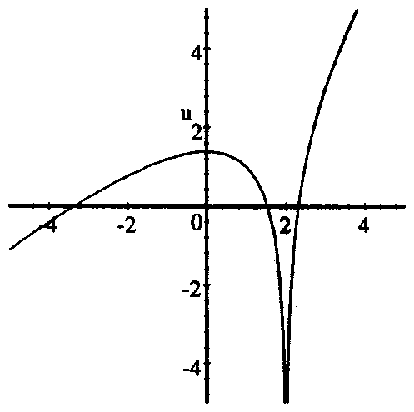}   } & {\includegraphics[
natheight=2.000300in,  natwidth=2.332400in, height=2.0003in, width=2.3324in
]   {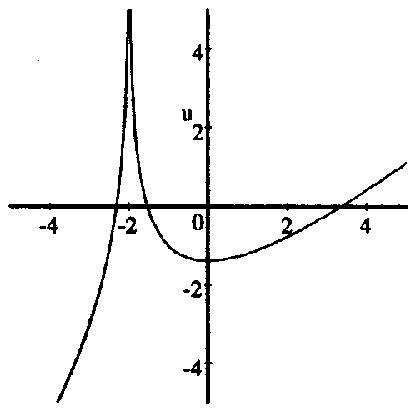}   }
\\
u=\beta+2\ln\left|  \beta-2\right|  & u=\beta-2\ln\left|
\beta+2\right|\\Fig. 1
\end{array}
\]
Thus, one can see that for $A\neq0$ there are up to three possibilities for
$\beta=\beta\left(  u\right)  $ that correspond to the intervals of
monotonicity of $u\left(  \beta\right)  $. For instance, for $A>0$ these
are  $\left]  -\infty,0\right[  $, $\left]  0,A\right[  $, and $\left]
A,\infty\right[  $. In these regions the metric (\ref{met}) is regular and
has  some singularities along the curves $\beta=0$ and $\beta-A=0$.
\begin{remark}
Solutions characterized by $A_{0},\,u_{0,}\,\beta_{0}\left( u_{0}\right) $
are globally diffeomorphic, \textit{via }$x\leftrightarrow-x$, to solutions
characterized by $A_{1}=-A_{0},\,u_{1}=-u_{0,\,}\beta_{1}\left(
u_{1}\right) =-\beta_{0}\left( -u_{1}\right) $.
\end{remark}

\begin{thm}
Any Ricci-flat $4$-metric admitting a normal Killing algebra isomorphic to
$  so\left( 3\right) $ or $so\left( 2,1\right) $ with bidimensional leaves,
is of the form (\ref{form}).
\end{thm}

\begin{proof}
For what concerns the algebras isomorphic to $so\left(  2,1\right)  $\ it
is sufficient to observe that solutions obtained in \cite{1SVV00} (section
\textit{5}) (the case $h_{22}\neq0$) are locally of that form. As for the
Killing algebras isomorphic to $so\left(  3\right)  $,\ it is a direct
consequence of the results of (\cite{1SVV00} section 7).
\end{proof}

In the case of normal Killing algebras isomorphic to $\mathcal{K}
\mathit{il}\left(  dx^{2}\pm dy^{2}\right)  $ consider Ricci-flat manifolds
$M$ of the form
\begin{equation}
M\mathcal{=W \Sigma},\qquad g=\pi_{1}^{\ast}\left(  g_{\left[  u\right]
}\right)  +\pi_{1}^{\ast}\left(  u\right)  \pi_{2}^{\ast}\left(  g_{\Sigma
}\right) \label{mrof}
\end{equation}
where $\left(  \Sigma,g_{\Sigma}\right)  $ is a flat bidimensional manifold
and
\[
g_{\left[  u\right]  }=\epsilon\frac{1}{\sqrt{u}}\left(
du^{2}-\mathbb{\zeta }^{2}dv^{2}\right)
\]
with $\epsilon=\pm1$.
\begin{thm}
Any Ricci-flat $4$-metric, admitting a normal Killing algebra isomorphic to
$  Kil\left( dx^{2}\pm dy^{2}\right) $, and with bidimensional Killing
leaves,  is either of the form (\ref{mrof}) or flat.
\end{thm}

\begin{proof}
In the case of a Killing algebra isomorphic to \textit{$\mathcal{K}
$\textit{il}}$\left(  dx^{2}-dy^{2}\right)  $ it is just an interpretation
of  propositions 10 and 13 of \cite{1SVV00}. For instance, proposition 13
deals  with the case $\mathcal{\zeta}^{2}=-1$. Since in our case $w$ is
constant, the  lower block of $\mathbf{M}_{C}\left(  g\right)  $ has
obviously the form  $\pi_{1}^{\ast}\left(  u\right)  \pi_{2}^{\ast}\left(
g_{\Sigma}\right)  $  with $u=D\varphi+B$. Even more, if $D\neq0$ the
metric given by $\mathbf{M}
_{C}\left(  g\right)  $ is manifestly of the form (\ref{mrof}). If $D=0$, then
the upper block of $\mathbf{M}_{C}\left(  g\right)  $ can be brought to the
form $\gamma\left(  d\varphi^{2}+d\psi^{2}\right)  $, where $\varphi$ and
$\psi$ are conjugated harmonic functions and
$\gamma=\varepsilon\sqrt{\left|  B\right|  }$. This shows that the whole
metric is flat.    A similar argument can be applied to the case $h_{22}=0$
of section \textit{6}  of \cite{1SVV00} to conclude the proof for the case
$\mathcal{K}
\mathit{il}\left(  dx^{2}+dy^{2}\right)  $.
\end{proof}

\subsubsection{\textit{Case II}}

In this case, a coordinate-free description of global solutions, obtained
in a local form in section \textit{5} of \cite{1SVV00}, is as follows. Let
$\left(  \mathcal{W},u\right)  $ be as before, and $w$ be a solution of the
equation $d\left(  u\,J^{\ast}\left(  dw\right)  \right)  =0$. Consider
the flat indefinite Euclidean plane $\left(
\mathbb{R}^{2},\,\,d\xi^{2}
-d\eta^{2}\right)  $ introduced at the end of section \textit{3} of
\cite{1SVV00}. Then the direct product $M=\mathcal{W }\mathbb{R}^{2}$ can
be supplied with the following Ricci-flat metric
\begin{equation}
g=\pi_{1}^{\ast}\left(  g_{\left[  u\right]  }\right) +\pi_{1}^{\ast}\left(
u\right)  \pi_{2}^{\ast}\left( d\xi^{2}-d\eta^{2}\right)  +\pi_{1}^{\ast
}\left(  uw\right)
\pi_{2}^{\ast}\left(  \left(  \frac{d\xi-d\eta}{\xi-\eta  }\right)
^{2}\right)  ,\label{BLA}
\end{equation}
where $\pi_{1}:$ $M=\mathcal{W }\mathbb{R}^{2}\rightarrow\mathcal{W}$ and
$\pi_{2}:M=\mathcal{W }\mathbb{R}^{2}\rightarrow\mathbb{R}^{2}$ are natural
projections and
\[
g_{\left[  u\right]  }=\epsilon\frac{1}{\sqrt{u}}\left(
du^{2}-\mathbb{\zeta}  dv^{2}\right)
\]
with $\epsilon=\pm1$.   In the above construction one can substitute the
quotient $\mathbb{R}^{2}/T$  for $\mathbb{R}^{2}$, where $T$ denotes the
discrete group acting on  $\mathbb{R}^{2}$ generated by a transformation of
the form $\left(  \xi  ,\eta\right)  \longrightarrow\left(
\xi+a,\eta+a\right)  $, $a\in\mathbb{R}$.    Let now $\left(
\mathcal{W},u\right)  $ be as before but $w$ is a
$\mathbb{\zeta}$-harmonic function on $\mathcal{W}$. Then $M=\mathcal{W
}\mathbb{R}^{2}$ carries the Ricci-flat metric
\begin{equation}
g=\epsilon_{1}\pi_{1}^{\ast}\left(  du^{2}-\mathbb{\zeta}^{2}dv^{2}\right)
+\epsilon_{2}\pi_{2}^{\ast}\left(  d\xi^{2}-d\eta^{2}\right) +\pi_{1}^{\ast
}\left(  w\right)  \pi_{2}^{\ast}\left(  \left(
\frac{d\xi-d\eta}{\xi-\eta  }\right)  ^{2}\right) \label{BLAB}
\end{equation}
with $\epsilon_{i}=\pm1$.
\begin{thm}
Any Ricci-flat 4-metric, admitting a non-extendible bidimensional
non-commutative Killing algebra, is either of the form (\ref{BLA}) or (\ref
{BLAB}) with Killing leaves of one of two types $\mathbb{R}^{2}$ or $
\mathbb{R}^{2}/T$.
\end{thm}

\begin{proof}
It follows directly from the local description of solutions given in
proposition 10 and 13 of \cite{1SVV00} and observations at the end of
section
\textit{3} of \cite{1SVV00}.
\end{proof}

\begin{remark}
Recall that geometric singularities of multivalued solutions are classified
by types corresponding to finite dimensional commutative $\mathbb{R}$
algebras (see \cite{Vi87}). As it was noticed in the previous section, any
global solution, described there in invariant terms, can be  obtained from
the model one (see next section) as its pullback via a $\Bbb{
\zeta}$-holomorphic map of a $\Bbb{\zeta}$-complex $\mathcal{W}$
$\ $to $\mathbb{R}_{\Bbb{\zeta}}$. Singularities of this map are
interpreted naturally as geometric singularities of multivalued solutions
of the  Einstein equations over $\mathbb{R}_{\Bbb{\zeta}}   $. These
singularities are either  of type $\mathbb{C}$ or of type $
\mathbb{R}^{2}\oplus\mathbb{R}^{2}$. This shows that multivalued solutions
of the Einstein equations admit such types of singularities and give
further illustration of the theory (see \cite{Vi87}).
\end{remark}

\begin{remark}
It is important to stress the existence of superposition laws for some
classes of metrics. For instance, if in Case I one fixes an isomorphism
class  of \textit{3}-dimensional Killing algebras, a $\Bbb{\zeta}$-complex
curve $
\mathcal{W}$, a fiber type and a constant $A$, then the
corresponding solutions are parametrized by  $\Bbb{\zeta}$-harmonic
functions $u$ on $
\mathcal{W}$ and, therefore, constitute a linear space. If in Case
II one fixes a $\Bbb{\zeta}$-harmonic functions $u$ on $\mathcal{W}$, then
the corresponding metrics are parametrized by  solutions \ $w$ of the
linear equation $d\left( uJ^{*}\left(  dw\right) \right) =0$. Conversely,
one can fix $w$ and vary $u$.
\end{remark}

\begin{remark}
Ricci flat metrics described in invariant terms in this section are
\textit{  semi-global }in the sense that they are not extendible along the
Killing leaves, but, generally, they are extendable along the  orthogonal
leaves. If the pair $\left( \mathcal{W},u\right) $,  over which a
Ricci-flat metric of the considered type is  constructed, is nonextendable,
then the metric itself is also  nonextendable. In this sense the results of
this section give all  global forms of the considered class of Ricci-flat
metrics.
\end{remark}

\section{Model solutions, normal coordinates and the local geometry of
leaves\label{ncgflgl}}   In view of proposition \textit{2} the basic
geometrical properties of the  metrics we are considering can be extracted
from the model solutions. This is  why they deserve special attention.
Recall that by a \textit{model solution} we understand one for which
$\left(
\mathcal{W},u\right)  =\left(  \mathbb{R}_{\mathbb{\zeta}}^{2},x\right)  $.
Thus, coordinates $\left(  x,y\right)  $, corresponding to the canonical
$\mathbb{\zeta}$-complex variable $x+\mathbb{\zeta} y$, are privileged in
$\mathbb{R}_{\mathbb{\zeta}}^{2}$ and also, \textit{via }the pullback $\pi
_{1}^{\ast}$, on orthogonal leaves. These coordinates will be used to
introduce \textit{normal coordinates} which are more convenient from a
geometric view-point.    It is worth to emphasize that
$\mathbb{\zeta}$-complex coordinates $\left(  x,y\right)  $, appearing
automatically in our approach, generalize the well  known Kruskal-Szekeres
coordinates for the Schwarzschild metric. By their  nature they do not
depend on a choice of the branch of the function  $(\beta-A)/\beta$.
Metrics, we are analyzing, may be regarded\ as the \textit{composition} of
two  families of bidimensional metrics, one along the Killing leaves and
another  along the orthogonal leaves. Then, the \textit{composed} metrics
can be  characterized in terms of intrinsic and extrinsic peculiarities of
these  bidimensional metrics.
\textbf{Killing Leaves}

\textit{They are homogeneous surfaces. Then, their intrinsic geometry is
completely characterized by the value of the Gauss curvature. This value is
a known function of }$u$\textit{\ (of }$x$\textit{\ for model solutions)
and it  remains to describe their extrinsic geometry, for instance their
geodesic curvature.}
\textit{The geodesic flow, which corresponds to a non-commutatively integrable
Hamiltonian vector field (proposition \ref{gfi}), projects on the geodesic
flow of the metric restricted to a Killing leave.}
\textbf{Orthogonal leaves}

\textit{They have a trivial extrinsic geometry, namely, they are totally
geodesic submanifolds (proposition \ref{gfi}). Thus, it remains to
characterize their intrinsic geometry. To this purpose, notice that these
leaves admit a Killing field, given\ in coordinates }$\left(  u,v\right)
$\textit{\ (}$\left(  x,y\right)  $\textit{\ for model solutions) by
}$\partial_{v}$\textit{. As it is well known, all scalar differential
invariants of such bidimensional metrics are functions of their Gauss
curvature }$K$\textit{. Moreover, the metric itself is completely
characterized by the function }$I=I\left(  K\right)  $\textit{\ for a
single  independent (for generic metrics) differential invariant
}$I$\textit{. For the  model solution this function is given for }$I=\left|
\nabla K\right|  ^{2}$\textit{.}    The behavior of geodesics, which is
particularly important from the physical  view-point, is described below
for metrics admitting a \textit{3}-dimensional  Killing algebra. Namely, we
show that the corresponding geodesic flows are
\textit{non-commutatively integrable}, by explicitly exhibiting all necessary
first integrals. The integrability of metrics with non-extendible Killing
algebras is at the moment unclear.    In what follows we report in a
systematic way all the aforementioned data for  the model solutions. Each
case is labelled by a suitable subset of the  following set of parameters
$\left\{  \mathcal{G},\mathbb{\zeta}
^{2},A,branch,\epsilon_{1},\epsilon_{2},\epsilon\right\}  $, where
$\epsilon_{1},$ $\epsilon_{2}=\pm1$ are the constants in the equation
(\ref{BLAB}) and $\epsilon=\pm1$ is linked to the Gauss curvature $K_{0}$
of  Killing leaves.    Note that in the last case model solutions of the
form (\ref{BLA}),  corresponding to a given label, are parametrized by
solutions $w$ of the  equation
\[
d\left(  u\,J^{\ast}\left(  dw\right)  \right)  =0,
\]
while those of the form (\ref{BLAB}) are parametrized by $\mathbb{\zeta}$
-harmonic functions on $\mathbb{R}_{\mathbb{\zeta}}^{2}$.

It is easy to see that metrics corresponding to different labels, or
different parametrizing functions are not \textit{isometric}.
\subsection{$\mathcal{G}=so\left(  2,1\right)  ,$ $\mathcal{G}=so(3)$}

In this case the model solutions are either flat metrics ( $A=0$) or have
the following local expression in terms of the canonical coordinates
$\left(  x,y\right)  $ on $\mathcal{W}$ (see (\ref{form}))
\begin{equation}
g=\epsilon_{1}\frac{\beta-A}{\beta}\left(  dx^{2}-\mathbb{\zeta}^{2}
dy^{2}\right)  +\epsilon_{2}\beta^{2}\left(  d\vartheta^{2}+\digamma\left(
\vartheta\right)  d\varphi^{2}\right) \label{m1}
\end{equation}
where $\digamma\left(  \vartheta\right)  $ is equal, in the case of
$so\left( 2,1\right)  $, either to $\sinh^{2}\vartheta$ or to
$-\cosh^{2}\vartheta$,  depending on the signature of the metric, and to
$\sin^{2}\vartheta$ in the  case of $so\left(  3\right)  $. The function
$\beta\left(  x\right)  $ is  defined by one of the positive branches of
the tortoise equation
\[
\beta+A\log\left|  \beta-A\right|  =x
\]
with $A\neq0$.   By introducing the \textit{normal coordinates} $\left(
r=\beta,\tau
=y,\vartheta,\varphi\right)  $, the local expression of $g$ becomes
\begin{equation}
g=\epsilon_{1}\left(  \frac{r}{r-A}dr^{2}-\mathbb{\zeta}^{2}\frac{r-A}{r}
d\tau^{2}\right)  +\epsilon_{2}r^{2}\left(  d\vartheta^{2}+\digamma\left(
\vartheta\right)  d\varphi^{2}\right) \label{rot}
\end{equation}

\subsubsection{Killing leaves}

\paragraph{The Gaussian curvature $K.$}

In the cases $so\left(  2,1\right)  $ or $so\left(  3\right)  $ it is given
by
\begin{align*}
K_{so\left(  2,1\right)  }  & =-\frac{\epsilon_{2}}{r^{2}}\\ K_{so\left(
3\right)  }  & =\frac{\epsilon_{2}}{r^{2}},
\end{align*}
respectively.   Thus, as already mentioned in section 3 of \cite{1SVV00},
we see that the  Killing leaves are bidimensional Riemannian manifolds,
with constant Gauss  curvature whose value, depending on $r$, changes with
the leaf.    In the case of $so\left(  2,1\right)  $ the Killing leaves are
\textit{non-Euclidean} or \textit{''anti'' non-Euclidean }planes, depending on
whether $\epsilon_{2}$ is positive or negative, assuming that
$\digamma\left(
\vartheta\right)  =\sinh^{2}\vartheta$, while they have an indefinite metric
with nonvanishing constant Gauss curvature if $\digamma\left(  \vartheta
\right)  =$ $-\cosh^{2}\vartheta$.

In the case of $so\left(  3\right)  $, the Killing leaves are standard
(metric) spheres or ''anti''-spheres, depending on whether $\epsilon_{2}$
is  positive or negative.
\paragraph{The second fundamental form $II$}

We evaluate its components with respect to normal unit vector fields
parallel to the coordinate fields, namely
\[
II\left(  X,Y\right)  =\left(  \nabla_{X}Y,n_{r}\right)  n_{r}+\left(
\nabla_{X}Y,n_{y}\right)  n_{y},
\]
where $X$ and $Y$ are tangent to the Killing leaves, and
$n_{i}=\sqrt{\left| g^{ii}\right|  }\partial_{i}$; here the index $i$ is
either $r$ or $y$. In the  normal coordinate the associated matrices
$II^{y}$and $II^{r}$ are
\begin{align*}
\left(  II_{ab}^{r}\right)   & =\sqrt{\left|  g_{rr}\right|  }\left(
\Gamma_{ab}^{r}\right)  =-\frac{\epsilon_{1}}{r}\left(  \frac{r-A}{r}\right)
\sqrt{\left|  \left(  \frac{r}{r-A}\right)  \right|  }\mathbf{H}\\
\left(  II_{ab}^{y}\right)   & =\sqrt{\left|  g_{yy}\right|  }\left(
\Gamma_{ab}^{y}\right)  =0,
\end{align*}
where $\mathbf{H}$ is the lower block of the matrix associated to $g$ in
the normal coordinates, and the $\Gamma$'s are the Christoffel symbols of
$g.$
\paragraph{The Christoffel symbols of the normal connection $\widetilde
{\Gamma}_{ai}^{j}$}   Since the normal coordinates are adapted coordinates,
the Christoffel symbols  are defined by
\[
\widetilde{\Gamma}_{ai}^{j}\partial_{j}=\nabla_{a}\left(  \partial_{i}\right)
^{\perp},
\]
so that
\[
\widetilde{\Gamma}_{ai}^{j}=\Gamma_{ai}^{j}=0.
\]

\paragraph{The geodesic curvature of Killing leaves}

Recall that the \textit{geodesic curvature} $C_{g}$ of a curve on $M$ whose
parametric equations are given by $x_{\mu}\left(  s\right)  $, where $s$ is
a
\textit{natural parameter},\textit{\ }is given by
\[
C_{g}^{2}=g_{\mu\nu}\Phi^{\mu}\Phi^{\nu}
\]
with $\Phi^{\lambda}\equiv\dot{V}^{\lambda}+\Gamma_{\mu\nu}^{\lambda}V^{\mu
}V^{\nu},$ where $V^{\mu}=\dot{x}_{\mu}$ are the components of the tangent
vector to the curve.    Observe that the curve, defined by $\varphi=0$ on a
Killing leaf, is a  geodesic for the metric restricted to a Killing leaf.
In normal coordinates  this geodesic $\gamma$, considered as a curve on
$M$, is given by
\[
r=\text{\textit{const.}},\quad\text{ }y=0,\quad\vartheta=\frac{s}{r}
,\quad\varphi=0
\]
so that
\[
V^{r}=0,\quad\text{ }V^{y}=0,\quad V^{\vartheta}=\frac{1}{r},\quad
V^{\varphi }=0,
\]
and
\[
\Phi^{\lambda}=\Gamma_{\vartheta\vartheta}^{\lambda}=\delta_{\lambda r}
\Gamma_{\vartheta\vartheta}^{r}=-\epsilon_{1}\delta_{\lambda r}\left(
r-A\right)  .
\]
Therefore, we obtain
\begin{equation}
C_{g}^{2}=\epsilon_{1}r\left(  r-A\right)  .\label{gc1}
\end{equation}
By obvious symmetry arguments the geodesic curvature of any geodesic on a
Killing leaf with respect to the induced metric is given by (\ref{gc1}).
So,  this quantity characterizes completely the geodesic curvature of the
Killing leaves.
\subsubsection{ Orthogonal leaves}

\paragraph{The scalar differential invariants}

The Gaussian curvature $K$ and $\left|  \nabla K\right|  ^{2}$are
\begin{align*}
K  & =\frac{\epsilon_{1}A}{r^{3}}\\
\left|  \nabla K\right|  ^{2}  & =9K^{3}\left(  1+\left(  \frac{1}{A^{2}
K}\right)  ^{\frac{1}{3}}\right)  .
\end{align*}

\paragraph{The Christoffel symbols of the normal connection $\widetilde
{\Gamma}_{ia}^{b}$}   They are given by
\[
\widetilde{\Gamma}_{ra}^{b}=\Gamma_{ra}^{b}=\delta_{ab}\frac{1}{r}
,\quad\widetilde{\Gamma}_{ya}^{b}=\Gamma_{ya}^{b}=0.
\]

\subsection{$\mathcal{G}$=\textit{$\mathcal{K}$\textit{il}}($d\xi^{2}\pm
d\eta^{2}$)}   In this case the model solutions are either flat metrics or
have the following  local expression in terms of the coordinates $\left(
x,y\right)  $ which in  this case are also normal coordinates (see
(\ref{mrof})):
\[
g=\epsilon_{1}\frac{1}{\sqrt{x}}\left(  dx^{2}-\mathbb{\zeta}^{2}
dy^{2}\right)  +\epsilon_{2}x\left(  d\xi^{2}+\epsilon d\eta^{2}\right)
\]
where $\epsilon_{i}=\pm1$.
\subsubsection{Killing leaves}

They are flat $2$-manifolds.
\paragraph{The second fundamental form $II$}

As before, we evaluate its components with respect to normal unit vector
fields parallel to the coordinate fields. The matrices $II^{x}$and $II^{y}$
are given by
\begin{align*}
\left(  II_{ab}^{x}\right)   & =\sqrt{\left|  g_{xx}\right|  }\left(
\Gamma_{ab}^{x}\right)  =-\frac{\epsilon_{1}}{2}\left|  x\right|  ^{-\frac
{3}{4}}\mathbf{H}\\ II_{ab}^{y}  & =\sqrt{\left|  g_{yy}\right|
}\Gamma_{ab}^{y}=0,
\end{align*}
where $\mathbf{H}$ is the lower block of the matrix associated to $g$ in
the normal coordinates, and $\Gamma$'s are the Christoffel symbols of $g$.

\paragraph{The Christoffel symbols of the normal connection}

They are given by
\[
\widetilde{\Gamma}_{ai}^{j}=\Gamma_{ai}^{j}=0
\]

\paragraph{The geodesic curvature of the Killing leaves}

As before, in the case of non null geodesics, it is characterized by
\[
C_{g}^{2}=\epsilon_{1}\frac{x^{-\frac{3}{4}}}{4}.
\]

\subsubsection{Orthogonal leaves}

\paragraph{The scalar differential invariants}

The Gaussian curvature $K$ and $\left|  \nabla K\right|  ^{2}$ are given by

\begin{align*}
K  & =-\frac{\epsilon_{1}}{4x^{\frac{3}{2}}}\\
\left|  \nabla K\right|  ^{2}  & =-9K^{3}
\end{align*}

\paragraph{The Christoffel symbols of the normal connection}

They are given by $\widetilde{\Gamma}_{ia}^{b}$. Thus, we have
\begin{align*}
\widetilde{\Gamma}_{\rho a}^{b}  & =\Gamma_{\rho a}^{b}=\frac{1}{2}\delta
_{ab}\\
\widetilde{\Gamma}_{\sigma a}^{b}  & =\Gamma_{\sigma a}^{b}=0.
\end{align*}

\subsection{Non-extendible bidimensional non-commutative Killing algebra}

In this case, the coordinates $\left(  x,y\right)  $ are also normal
coordinates, and the model solutions have one of the following local
expressions (see (\ref{BLA}), (\ref{BLAB})):
\begin{equation}
g=\epsilon_{1}\frac{1}{\sqrt{x}}\left(  dx^{2}-\mathbb{\zeta}^{2}
dy^{2}\right)  +\epsilon_{2}x\left[  \left(  d\xi^{2}-d\eta^{2}\right)
+w\left(  \frac{d\xi-d\eta}{\xi-\eta}\right)  ^{2}\right]  ,\label{gx}
\end{equation}
where $w$ is a solution of the equation
\[
x\left(  \frac{\partial^{2}}{\partial x^{2}}-\mathbb{\zeta}^{2}\frac
{\partial^{2}}{\partial y^{2}}\right)  w+\frac{\partial w}{\partial x}=0;
\]
or
\begin{equation}
g=\epsilon_{1}\left(  dx^{2}-\mathbb{\zeta}^{2}dy^{2}\right)  +\epsilon
_{2}\left[  \left(  d\xi^{2}-d\eta^{2}\right)  +w\left(  \frac{d\xi-d\eta}
{\xi-\eta}\right)  ^{2}\right] \label{g1}
\end{equation}
where $w$ is a $\mathbb{\zeta}$-harmonic function.
\subsubsection{Killing leaves}

They are flat manifolds.
\paragraph{The second fundamental form $II$}

As before, we evaluate its components with respect to normal unit vector
fields parallel to the coordinate fields. The matrices $II^{x}$ and
$II^{y}$  for the metric (\ref{gx}), are given by
\begin{align*}
\left(  II_{ab}^{x}\right)   & =\sqrt{\left|  g_{xx}\right|  }\left(
\Gamma_{ab}^{x}\right)  =-\frac{\epsilon_{1}\epsilon_{2}}{2}\left|  x\right|
^{-\frac{3}{4}}\left[  \left(  g_{ab}\right)  +x^{2}\frac{\partial_{x}
w}{\left(  \xi-\eta\right)  ^{2}}\left(
\begin{array}
[c]{cc}  1 & -1\\
-1 & 1
\end{array}
\right)  \right] \\
\left(  II_{ab}^{y}\right)   & =\sqrt{\left|  g_{yy}\right|  }\left(
\Gamma_{ab}^{y}\right)  =\frac{\epsilon_{1}\epsilon_{2}\mathbb{\zeta}^{2}}
{2}\left|  x\right|  ^{\frac{5}{4}}\frac{\partial_{y}w}{\left(  \xi
-\eta\right)  ^{2}}\left(
\begin{array}
[c]{cc}  1 & -1\\
-1 & 1
\end{array}
\right)  ,
\end{align*}
while for the metric (\ref{g1}) one has
\begin{align*}
\left(  II_{ab}^{x}\right)   & =\sqrt{\left|  g_{xx}\right|  }\left(
\Gamma_{ab}^{x}\right)  =-\frac{\epsilon_{1}\epsilon_{2}}{2}\frac{\partial
_{x}w}{\left(  \xi-\eta\right)  ^{2}}\left(
\begin{array}
[c]{cc}  1 & -1\\
-1 & 1
\end{array}
\right) \\
\left(  II_{ab}^{y}\right)   & =\sqrt{\left|  g_{yy}\right|  }\left(
\Gamma_{ab}^{y}\right)  =\frac{\epsilon_{1}\epsilon_{2}\mathbb{\zeta}^{2}}
{2}\frac{\partial_{y}w}{\left(  \xi-\eta\right)  ^{2}}\left(
\xi-\eta\right)
^{2}\left(
\begin{array}
[c]{cc}  1 & -1\\
-1 & 1
\end{array}
\right)  ,
\end{align*}

\paragraph{The Christoffel symbols of the normal connection:}

\[
\widetilde{\Gamma}_{ai}^{j}=\Gamma_{ai}^{j}=0
\]

\paragraph{The geodesic curvature of Killing leaves.}

The geodesic curvature for the metric (\ref{gx}) is given by
\[
C_{g}^{2}=\frac{\epsilon_{1}\left|  x\right|  ^{-\frac{3}{2}}}{4}\left[
\left(  1+xV\partial_{x}w\right)  ^{2}-\left(  xV\partial_{y}w\right)
^{2}\right]  ,
\]
where $V=\epsilon_{2}\left(  \frac{x}{\left(  \xi-\eta\right)  ^{2}}\right)
\left(  V^{\xi}-V^{\eta}\right)  ^{2}$ with $V^{\xi}$ and $V^{\eta}$
components of a vector tangent to a geodesic of the restricted metric.
Their expression is
\begin{align*}
V^{\xi}  & =\frac{1}{2}\frac{\left(  a+k\right)  \left(  \xi-\eta\right)
-3kw}{\sqrt{\left|  kx\left[  a\left(  \xi-\eta\right)  ^{2}-2wk\right]
\right|  }}\\
V^{\eta}  & =\frac{1}{2}\frac{\left(  a-k\right)  \left(  \xi-\eta\right)
-3kw}{\sqrt{\left|  kx\left[  a\left(  \xi-\eta\right)  ^{2}-2wk\right]
\right|  }}
\end{align*}
with $k$ and $a$ arbitrary constants.   For the metric (\ref{rot}) the
geodesic curvature is
\[
C_{g}^{2}=\frac{\epsilon_{1}}{4}\left[  1+\left(  \frac{V^{\xi}-V^{\eta}}
{\xi-\eta}\right)  ^{4}\left[  \left(  \partial_{x}w\right)  ^{2}-\left(
\partial_{y}w\right)  ^{2}\right]  +2\epsilon_{2}\left(  \frac{V^{\xi}
-V^{\eta}}{\xi-\eta}\right)  ^{2}\partial_{x}w\right]
\]

\subsubsection{Orthogonal leaves}

\paragraph{The scalar differential invariants.}

The Gaussian curvature $K$ and $\left|  \nabla K\right|  ^{2}$ are given by

\begin{gather*}
K=-\frac{\epsilon_{1}}{4x^{\frac{3}{2}}}\\
\left|  \nabla K\right|  ^{2}=-9K^{3}
\end{gather*}

\paragraph{The second fundamental form $II$}

As before, we evaluate its components with respect to normal unit vector
fields parallel to the coordinate fields, so that
\[
II_{ij}^{a}=\Gamma_{ij}^{a}\sqrt{\left|  g_{aa}\right|  }=0.
\]

\paragraph{The Christoffel symbols of the normal connection $\widetilde
{\Gamma}_{ia}^{b}$}   For the metric (\ref{gx}) they are
\begin{align*}
\left(  \widetilde{\Gamma}_{xa}^{b}\right)   & =\left(  \Gamma_{xa}
^{b}\right)  =\frac{1}{2x}\left(  \delta_{ba}\right)  +\frac{\epsilon_{2}}
{2}\frac{\partial_{x}\widetilde{w}}{\left(  \xi-\eta\right)  ^{2}}\left(
\begin{array}
[c]{cc}  1 & -1\\  1 & -1
\end{array}
\right) \\
\left(  \widetilde{\Gamma}_{ya}^{b}\right)   & =\left(  \Gamma_{ya}
^{b}\right)  =\frac{\epsilon_{2}}{2}\frac{\partial_{y}\widetilde{w}}{\left(
\xi-\eta\right)  ^{2}}\left(
\begin{array}
[c]{cc}  1 & -1\\  1 & -1
\end{array}
\right)  ,
\end{align*}
while for the metric (\ref{g1})
\[
\left(  \widetilde{\Gamma}_{ia}^{b}\right)  =\left(  \Gamma_{ia}^{b}\right)
=\frac{\epsilon_{2}}{2}\frac{\partial_{i}\widetilde{w}}{\left(  \xi
-\eta\right)  ^{2}}\left(
\begin{array}
[c]{cc}  1 & -1\\  1 & -1
\end{array}
\right)  ,\qquad i=x,y.
\]
In the above matrices $b$ is the row index and $a$ the column index:
\subsection{Geodesic flows}

Geodesic flows corresponding to the metrics we are dealing with show some
interesting properties, one of which will be used in next section when
describing the info-hole phenomenon. Below, having this application in
mind,  we shall discuss briefly only the flows, corresponding to the
metrics,  admitting a 3-dimensional Killing algebra.    Recall that the
geodesic flow, corresponding to a metric $g=g_{ij}dx_{i}   dx_{j}$ on
$\mathcal{M}$, can be viewed as the flow, generated by the  Hamiltonian
field $X_{H},\quad H=\frac{1}{4}g^{ij}p_{i}p_{j}$ ($p_{i}$'s are conjugated
to $x_{i}$'s coordinates on $T^{\ast}\mathcal{M}$). For example, up  to the
factor $\frac{1}{4}$, the Hamiltonian
\begin{equation}
\mathcal{H}=\epsilon_{1}\left[  \left(  1-\frac{A}{r}\right)  p_{r}
^{2}-\mathbb{\zeta}^{2}\frac{p_{\tau}^{2}}{1-\frac{A}{r}}\right]
+\frac{\epsilon_{2}}{r^{2}}\left(  p_{\vartheta}^{2}+G\left(  \vartheta
\right)  p_{\varphi}^{2}\right)  ,\label{gf1}
\end{equation}
corresponds to the metric (\ref{gc1}) expressed in the normal coordinates
(see subsection 3.1). Notice also that the projection
$\pi_{2}:M=\mathcal{W}    \Sigma\rightarrow\Sigma$ generates canonically a
projection
\[
\bar{\pi}_{2}:T^{\ast}M\rightarrow T^{\ast}\Sigma.
\]

\begin{proposition}
Projection $\pi_{2}$ sends geodesic (nonparametrized) curves of a model
Ricci-flat metrics associated with a $3$-dimensional Killing algebra to
geodesic (nonparametrized) curves of the metric $\left. g\right.
_{\Sigma}$.
\end{proposition}

\begin{proof}
It follows from the relation
\[
d_{a}\bar{\pi}_{2}\left(  X_{H,a}\right)  =\lambda X_{\left.  H\right.
_{\Sigma},\bar{\pi}_{2}\left(  a\right)  }\qquad\lambda\in\mathbb{R},\,a\in
T^{\ast}M,
\]
which, in its turn, is a direct consequence of the particular form of the
Hamiltonian (\ref{gf1}) (and similarly for algebras $Kil\left(  dx^{2}
+dy^{2}\right)  $.
\end{proof}

It is worth to note that geodesic flows, corresponding to the model
Ricci-flat metrics, associated with $3$-dimensional Killing algebras, are
integrable. To  see that one may observe that, for instance, Hamiltonian
(\ref{gf1}) with  $G\left(  \vartheta\right)
=\frac{1}{\sinh^{2}\vartheta}$ possesses five  independent first integrals
\[
\mathcal{H},\quad p_{\tau},\quad p_{\vartheta}^{2}+\frac{1}{\sinh^{2}
\vartheta}p_{\varphi}^{2},\quad\mathcal{I}_{1},\quad\mathcal{I}_{2}\quad
\]
where $\mathcal{I}_{1}$ and $\mathcal{I}_{2}$ are generators of a
noncommutative bidimensional subalgebra of $so\left(  2,1\right)  $, for
example
\begin{align*}
\mathcal{I}_{1}  & =\left[  \left(  1+\sqrt{2}\right)  \cos\varphi+\sin
\varphi\right]  p_{\vartheta}+\left[  1+\sqrt{2}+\coth\vartheta\left(
\cos\varphi-\left(  1+\sqrt{2}\right)  \sin\varphi\right)  \right]
p_{\varphi}\\
\mathcal{I}_{2}  & =\sqrt{2}\left[  \cos\varphi+\sin\varphi\right]
p_{\vartheta}+\left[  2+\sqrt{2}\coth\vartheta\left(  \cos\varphi-\sin
\varphi\right)  \right]  p_{\varphi}.
\end{align*}
The $5$ first integrals span a rank $3$ Lie algebra $\mathcal{A}$, and
since
\[
\operatorname{rank}\mathcal{A}+\dim\mathcal{A}=\dim T^{\ast}M\text{\quad and
\quad}\operatorname{rank}\mathcal{A}\text{ }<\dim\mathcal{A}
\]
the system is noncommutatively integrable in the sense \cite{MF78, SV00}.
Note also that the above proposition is no more valid for the geodesic flow
of  Ricci-flat metrics with nonextendable bidimensional Killing algebras.
\subsection{The geodesic flow (non-extendible case)}

Here, apart from the solution $w=$ \textit{const}, the geodesic curves do
not project on the Killing fields and $h^{ab}p_{a}p_{b}$ is no more a first
integral for the geodesic equations.
\section{Examples\label{esempi}}

In this concluding section we illustrate the previous general results with
a few examples.    According to proposition \textit{\ref{pb}}, we can
construct any solution as  the pullback of a model solution \textit{via} a
$\zeta$-holomorphic map $\Phi$  of a $\zeta$-complex curve $\mathcal{W}$ to
$\mathbb{R}_{\mathbb{\zeta}}^{2}$.  Recall that in the pair $\left(
\mathcal{W},u\right)  $, describing the so  obtained solution,
$u=\operatorname{Re}$ $\Phi$.    A detailed analysis of geometrical
properties of the obtained exact solutions,  as well as their possible
physical interpretation, is postponed up in a  forthcoming paper.
\bigskip

\subsection{\textit{Algebraic solutions}}

Let $\mathcal{W}$ be an algebraic curve over $\mathbb{C}$, understood as a
$\zeta$-complex curve with $\zeta^{2}=-1$.    With a given meromorphic
function $\Phi$ on $\mathcal{W}$ a pair $\left(
\mathcal{W}_{\Phi},u\right) $ is associated, where $\mathcal{W}_{\Phi}$ is
$\mathcal{W}$ deprived of the poles of $\Phi$ and $u$ the real part of
$\Phi$.   A solution (metric) constructed over such a pair will be called
\textit{algebraic. }Algebraic metrics are generally singular. For instance,
such a metric is degenerate along the fiber $\pi_{1}^{-1}\left(  a\right) $
(see section \textit{\ref{global}}) if $a\in\mathcal{W}$ is such that
$d_{a}u=0$.
\subsection{\textit{Info-holes}}

Space-times corresponding to algebraic metrics and, generally, to metrics
with signature equal to $2$ constructed over complex curves
($\zeta^{2}=-1$),  exhibit the following interesting property: \textit{for
a given observer there  exists another observer which can be never
contacted}.    By defining an \textit{info-hole
}(information-hole)\textit{\ }of a given  point $a$ to be the set of points
of the space-time whose future does not  intersect the future of $a$, the
above property can be paraphrased by saying  that \textit{the info-hole of
a given point of such a space-time is not empty}.    In fact, consider a
metric of the form (\ref{form}) constructed over a complex  curve whose
standard fiber $\Sigma$ is a bidimensional manifold supplied with  an
indefinite metric of constant Gauss curvature equal to $1$. For our
purpose, it is convenient to take for $\Sigma$ the hyperboloid $x_{1}
^{2}+x_{2}^{2}-x_{3}^{2}=1$, supplied with the induced metric $\left.
g\right|  _{\Sigma}=\left.  dx_{1}^{2}+dx_{2}^{2}-dx_{3}^{2}\right|
_{\Sigma} $.\ The \textit{light-cone }of $\left.  g\right|  _{\Sigma}$ at a
given point  $b$ is formed by the pair of rectilinear generators of
$\Sigma$ passing  through $b$. Since the geodesics of $g$ project
\textit{via }$\pi_{2}$ into  geodesics of $\left.  g\right|  _{\Sigma}$
(see section \textit{\ref{ncgflgl}   }), it is sufficient to prove the
existence of info-holes for the  bidimensional space-time $\left(
\Sigma,\left.  g\right|  _{\Sigma}\right)  $. To this purpose, consider the
standard projection $\pi$ of $\mathbb{R}
^{3}=\left\{  \left(  x_{1},x_{2},x_{3}\right)  \right\}  $ onto
$\mathbb{R}^{2}=\left\{  \left(  x_{1},x_{2}\right)  \right\}  $. Then,
$\left.  \pi\right|  _{\Sigma}$ projects $\Sigma$ onto the region $x_{1}
^{2}+x_{2}^{2}\geq1$ in $\mathbb{R}^{2}$ and the rectilinear generators of
$\Sigma$ are projected onto tangents to the circle $x_{1}^{2}+x_{2}^{2}=1$.
Suppose that the time arrow on $\Sigma$ is oriented according to increasing
value of $x_{3}$. Then the future region $F\left(  b\right)
\subset\Sigma$ of  the point $b\in\Sigma$, $b\left(  1,1,\beta\right)  \
\beta>0$, projects onto  the domain defined by
$D=\{x\in\mathbb{R}^{2}:x_{1}>1,x_{2}>1\}$ , and the  future region of any
point $b^{\prime}\in\Sigma$, such that $\pi\left(  b^{\prime}\right)  \in
D^{\prime}$ and $x_{3}\left(  b^{\prime}\right)  >0$,  does not intersect
$F\left(  b\right)  $. By obvious symmetry arguments the  result is valid
for any point $b\in\Sigma$
\[
\begin{array}
[c]{c}
\
\raisebox{--0.0173in}{\includegraphics[
 natheight=2.008100in,  natwidth=2.408500in,
height=2.0081in,  width=2.4085in  ]   {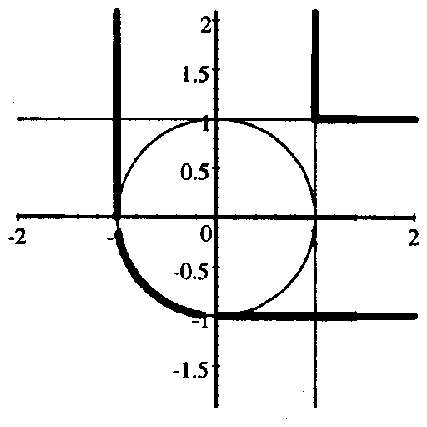}   }
\\Fig. 2
\end{array}
\]
\ \
\subsection{\textit{A star ''outside'' the universe}}

The Schwarzschild solution shows a ''star'' generating a space ''around''
itself. It is an $so\left(  3\right)  $-invariant solution of the vacuum
Einstein equations. On the contrary, its $so\left(  2,1\right)  $-analogue
(see section \textit{\ref{ncgflgl}) shows a ''star'' generating the space
only  on ''one side of itself''. More exactly, the fact that the space in
the  Schwarzschild universe is formed by a 1-parametric family of
''concentric''  spheres allows one to give a sense to the adverb
''around''. In the  }$so\left(  2,1\right)  $-case the space is formed by a
1-parameter family of  ''concentric'' hyperboloids. The adjective
''concentric'' means that the  curves orthogonal to hyperboloids are
geodesics and metrically converge to a  singular point. This explains in
what sense this singular point generates the  space only on ''one side of
itself''.
\subsection{\textit{Kruskal-Szekeres type solutions}}

We describe now a family of solutions which are of the
\textit{Kruskal-Szekeres type }\cite{Wa84}, namely, that are characterized as
being maximal extensions of the local solutions determined by an affine
parametrization of null geodesics, and also by the use of more than one
interval of monotonicity of $u\left(  \beta\right)  $.    Consider the
$\zeta$-complex curve
\[
\mathcal{W}=\left\{  \left(  z=x+\zeta y\right)  \in\mathbb{R}_{\zeta}
^{2}:y^{2}-x^{2}<1\right\}  ,\,\,\,\zeta^{2}=1
\]
and the $\zeta$-holomorphic function $\Phi:\mathcal{W}\rightarrow
\mathbb{R}_{\zeta}^{2}$
\[
\Phi\left(  z\right)  =A\ln\left(  \left|  A\right|  z^{2}\right)  =A\left(
\ln\left|  A\left(  x^{2}-y^{2}\right)  \right|  +\mathbb{\zeta}\ln\left|
\frac{x+y}{x-y}\right|  \right)  .
\]
Thus, in the pair $\left(  \mathcal{W},u\right)  $ the $\mathbb{\zeta}
$-harmonic function $u$ is given by
\[
u=A\ln\left|  A\left(  x^{2}-y^{2}\right)  \right|  .
\]
Let us decompose $\mathcal{W}$ in the following way:
\[
\mathcal{W=U}_{1}\cup\mathcal{U}_{2}
\]
where
\begin{align*}
\mathcal{U}_{1}  & =\left\{  \left(  z=x+\zeta y\right)  \in\mathbb{R}_{\zeta
}^{2}:0\leq y^{2}-x^{2}<1\right\} \\
\mathcal{U}_{2}  & =\left\{  \left(  z=x+\zeta y\right)  \in\mathbb{R}_{\zeta
}^{2}:y^{2}-x^{2}\leq0\right\}  .
\end{align*}
Consider now the solution defined as the pull back with respect to $\left.
\Phi\right|  _{\mathcal{U}_{1}}$ and $\left.  \Phi\right|  _{\mathcal{U}_{2}}$
of the model solutions determined by the following data: in the case of
$\left.  \Phi\right|  _{\mathcal{U}_{1}}$, $\mathcal{G}=so\left(  3\right)
$  or $\mathcal{G}=so\left(  2,1\right)  $, (see equation (\ref{m1})),
characterized by $\digamma\left(  \vartheta\right)  =sin^{2}\vartheta$ or
$\digamma\left(  \vartheta\right)  =sinh^{2}\vartheta$ respectively,
$\epsilon_{1}=\epsilon_{2}=1$, $A>0$, and for $\beta\left(  u\right)  $ the
interval $\left]  0,A\right]  $; in the case of $\left.  \Phi\right|
_{\mathcal{U}_{2}}$ the same data except for $\beta\left(  u\right)  $ which
belongs to the interval $\left[  A,\infty\right[  $. The case
$\digamma\left(
\vartheta\right)  =sin^{2}\vartheta$, corresponding to $so\left(  3\right)  $,
will give the Kruskal-Szekeres solution (see \cite{Wa84}). The case
$\digamma\left(  \vartheta\right)  =sinh^{2}\vartheta$, corresponding to
$so(2,1)$, will differ from the previous one in the geometry of the Killing
leaves, which will now have a negative constant Gaussian curvature. The
metric  $g$ has the following local form
\[
g=4A^{3}\frac{\exp\frac{\beta}{A}}{\beta}\left(  dy^{2}-dx^{2}\right)
+\beta^{2}\left[  d\vartheta^{2}+\digamma\left(  \vartheta\right)
d\varphi^{2}\right]  ,
\]
the singularity $\beta=0$ occurring at $y^{2}-x^{2}=1$.
\subsection{\textit{The ''square root'' of the Schwarzschild universe}}

Now we will discuss Einstein metrics of signature $2$, induced by a $\zeta
$\textit{-quadratic} map, $\zeta^{2}=1$, from a model of the Schwarzschild
type. The Einstein manifolds obtained in this way are interpreted naturally
as
\textit{parallel universes }generated by\textit{\ } \textit{parallel ''stars''.}

$\mathbf{1.}$ Further on it is assumed that $\zeta^{2}=1$. Recall that a
model metric of Schwarzschild type is characterized by the following data:
\[
\mathcal{G}=so\left(  3\right)  \text{ or }\mathcal{G}=so\left(  2,1\right)
,\text{ }\epsilon_{1}=\epsilon_{2}=1,\text{ }A>0,\text{ }\beta\left(
u\right)  \in\left[  A,\infty\right[
\]
Its \textit{square root} is the metric induced from it by the $\zeta
$\textit{-quadratic} map $z\longmapsto\frac{\mathbb{\zeta}}{2}z^{2}$ of
$\mathbb{R}_{\mathbb{\zeta}}^{2}$ into itself. Obviously, the basic pair
$\left(  \mathcal{W},u\right)  $ determining the square root is $\left(
\mathbb{R}_{\mathbb{\zeta}}^{2},xy\right)  ,\quad z=x+\mathbb{\zeta} y$,

The local expression of the metric is
\begin{equation}
g=-\frac{1}{2}\frac{\exp\frac{\beta+xy}{A}}{\beta}\left(
y^{2}-x^{2}\right)
\left(  dy^{2}-dx^{2}\right)  +\beta^{2}\left[  d\vartheta^{2}+\digamma\left(
\vartheta\right)  d\varphi^{2}\right]
\end{equation}
where $\digamma$, depending on the Gauss curvature $K$, is one of the
functions $\sin^{2}\vartheta$, $\sinh^{2}\vartheta$, $-\cosh^{2}\vartheta$
according to equation (\ref{m1}). The metric is degenerate along the lines
$y=\pm x$. For our purposes it is convenient to refer the \textit{square
root  }metric to the Kruskal-Szekeres type coordinates considered in the
previous  example. To this end consider the holomorphic map
\[
\mathcal{F}:\mathbb{R}_{\mathbb{\zeta}}^{2}\longrightarrow\mathbb{R}
_{\mathbb{\zeta}}^{2}
\]
\[
\mathcal{F}\left(  z\right)  =\frac{1}{\sqrt{\left|  A\right|  }}\exp
\frac{\mathbb{\zeta} z^{2}}{4A}.
\]
Denote the quadrants of the $\zeta$-complex line $z=x+\zeta y$ bounded by
the lines $x=\pm y$ as
\[
I=\left\{  x\geq\left|  y\right|  \right\}  ,\quad II=\left\{  y\geq\left|
x\right|  \right\}  ,\quad III=\left\{  -x\geq\left|  y\right|  \right\}
,\quad IV=\left\{  -y\geq\left|  x\right|  \right\}  .
\]
$\mathcal{F}$ maps each of them onto the strip in the $\zeta$-complex line
$z^{\prime}=x^{\prime}+\zeta y^{\prime}$ enclosed by the thick lines as it
is  shown in the figure for $A=1$.
\[
\begin{array}
[c]{c}
\
\raisebox{--0.0173in}{\includegraphics[
natheight=2.008100in, natwidth=2.408500in,  height=2.0081in, width=2.4085in
]   {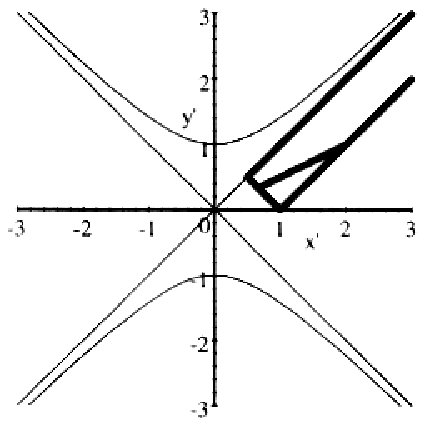}   }
\\
x^{\prime},y^{\prime}\\ Fig. 3
\end{array}
\]
\ \

The image of the lines $x=\pm y$ are respectively the half line $a$ and the
segment $b$. The third side of the strip corresponds to $x-y=\pm\infty$. In
\textit{normal coordinates} $r=\beta,\quad\tau=x^{2}+y^{2}$, a fixed value of
$\tau$\textit{\ }determines a\textit{\ }segment in each region ( the thick
segment $c$ in the figure). Its pull back with respect to $\pi_{1} $ (see
section \ref{gps}) is a spherical or hyperbolic shell for $\mathcal{G}=$
$so\left(  3\right)  $ or $\mathcal{G}=$ $so\left(  2,1\right)  ,$
respectively, (see example 3). Since the Einstein metrics we are
considering  is a pull back of a Schwarzschild type one it comes natural to
interpret the  maximal radius of each of these shells as the dimension of
the universe at the  corresponding instant of time, while the minimal
radius gives the instant  value of the Schwarzschild horizon.    The shells
corresponding to quadrants $I$ and $II$ are soldered along their  maximal
radius spheres (respectively, hyperboloids) forming one, say
$\mathcal{U}_{1}$, of two \textit{parallel universes.} Another one,
$\mathcal{U}_{2}$, is related similarly with quadrants $III$ and $IV$.
These  two universes have two common Schwarzschild horizons along which the
shells  corresponding to quadrants $II$, $III$ and $I$, IV , respectively,
are  soldered. Associating a ''star'' with each of these two horizons one
discovers  a system of two parallel stars in a perfect equilibrium, which
generate two
\textit{parallel universes,} $\mathcal{U}_{1}$ and $\mathcal{U}_{2}$ whose
dimensions grow infinitely with time.   $\mathbf{2.}$ As in example $4$ the
metric we are considering can be extended  beyond the Schwarzschild horizon
$x=y$. To this end consider
\begin{align*}
\mathcal{W}_{1}  & :=\left\{  z=x+\mathbb{\zeta} y\in\mathbb{R}_{\mathbb{\zeta
}}^{2}:\left|  x\right|  <\left|  y\right|  \Longrightarrow\,\,\,yx<A\ln
A\right\}  ,\,\,\,\,\zeta^{2}=1\\
\Phi\left(  z\right)   & =\frac{\mathbb{\zeta}}{2}z^{2}=xy+\frac{1}
{2}\mathbb{\zeta}\left(  x^{2}+y^{2}\right)  .
\end{align*}
Thus, in the basic pair $\left(  \mathcal{W}_{1},u\right)  $ the
$\mathbb{\zeta}$-harmonic function $u$ is given by
\[
u=xy.
\]
Let us decompose $\mathcal{W}_{1}$ in the following way:
\[
\mathcal{W}_{1}\mathcal{=V}_{1}\cup\mathcal{V}_{2}
\]
where
\begin{align*}
\mathcal{V}_{1}  & =\left\{  \left(  z=x+\zeta y\right)  \in\mathbb{R}_{\zeta
}^{2}:\left|  x\right|  \leq\left|  y\right|  ,\,\,\,yx<A\ln A\right\} \\
\mathcal{V}_{2}  & =\left\{  \left(  z=x+\zeta y\right)  \in\mathbb{R}_{\zeta
}^{2}:\left|  x\right|  \geq\left|  y\right|  \right\}  .
\end{align*}
Consider now the solution defined as the pull back with respect to $\left.
\Phi\right|  _{\mathcal{U}_{1}}$ and $\left.  \Phi\right|  _{\mathcal{U}_{2}}$
of the model solutions determined by the following data: in the case of
$\left.  \Phi\right|  _{\mathcal{V}_{1}}$
\[
\mathcal{G}=so\left(  3\right)  \text{ or }\mathcal{G}=so\left(  2,1\right)
,\text{ }\epsilon_{1}=\epsilon_{2}=1,\text{ }A>1,\text{ }\beta\left(
u\right)  \in\left]  0,A\right]  ,
\]
in the case of $\left.  \Phi\right|  _{\mathcal{V}_{2}}$
\[
\mathcal{G}=so\left(  3\right)  \text{ or }\mathcal{G}=so\left(  2,1\right)
,\text{ }\epsilon_{1}=\epsilon_{2}=1,\text{ }A>1,\text{ }\beta\left(
u\right)  \in\left[  A,\infty\right[  .
\]
In this case a branch change occurs along the singular lines $y=\pm x$
which become also \textit{first-type discontinuity} lines for $\beta$. In
the above  local expression there is a singularity at $\beta=0$, which is
also a  divergence of the scalar curvature.    Even in this case it is
convenient to refer the metric to the Kruskal-Szekeres  type coordinates.
In the case $\left|  x\right|  >\left|  y\right|  $ consider the
holomorphic map
\[
z^{\prime}=\frac{1}{\sqrt{\left|  A\right|  }}\exp\frac{\mathbb{\zeta}
z^{2}  }{4A}\qquad z^{\prime}=x^{\prime}+\mathbb{\zeta} y^{\prime}
\]
or,
\[
\left\{
\begin{array}
[c]{c}  U=y^{\prime}-x^{\prime}=-\frac{1}{\sqrt{\left|  A\right|
}}\exp-\frac{1}   {4A}\left(  y-x\right)  ^{2}\\
V=y^{\prime}+x^{\prime}=\,\,\,\frac{1}{\sqrt{\left|  A\right|  }}\,\exp
\frac{1}{4A}\left(  y+x\right)  ^{2},
\end{array}
\right.
\]
while, for $\left|  x\right|  <\left|  y\right|  $ and $yx<A\ln A$ consider
the map
\[
z^{\prime}=\frac{\mathbb{\zeta}}{\sqrt{\left|  A\right|  }}\exp\frac
{\mathbb{\zeta} z^{2}}{4A}
\]
or,
\[
\left\{
\begin{array}
[c]{c}  U=y^{\prime}-x^{\prime}=\frac{1}{\sqrt{\left|  A\right|
}}\exp-\frac{1}   {4A}\left(  y-x\right)  ^{2}\\
V=y^{\prime}+x^{\prime}=\,\,\,\frac{1}{\sqrt{\left|  A\right|  }}\,\exp
\frac{1}{4A}\left(  y+x\right)  ^{2},
\end{array}
\right.
\]
\[
\begin{array}
[c]{cc} {\includegraphics[ natheight=2.008100in,  natwidth=2.408500in,
height=2.0081in,  width=2.4085in  ]   {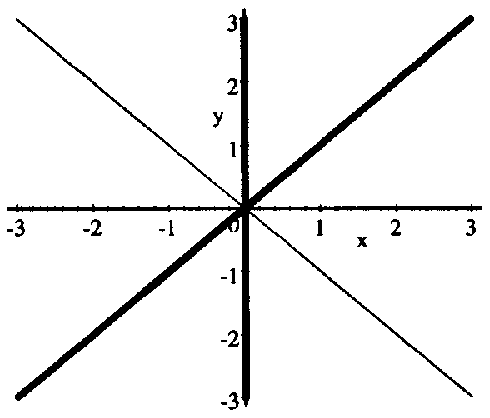}   } &
{\includegraphics[ natheight=2.008100in, natwidth=2.408500in,
height=2.0081in, width=2.4085in ]   {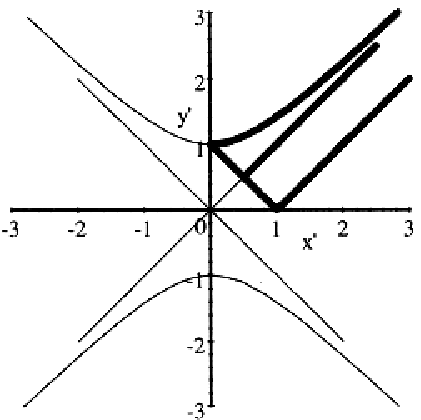}   }
\\
x,y & x^{\prime},y^{\prime}\\Fig. 4
\end{array}
\]

From the above figures, which correspond to the map with $A=1$, one can see
that the region defined by $\left|  x\right|  >$ $\left|  y\right|  $ is
mapped onto the lower strip. The region defined by $\left|  x\right|
<\left|  y\right|  $ and $yx<A\ln A$ is mapped onto the upper strip and
thus has the  singularity at $\beta=0$ as part of its boundary. For this
region the line  $x=y$, on which in the previous example occurred the
soldering of the shells  ($\tau=$ const.) along the maximal radius, now
lies beyond the singularity.  The lines $x-y=\pm\infty$ are mapped into the
line $x^{\prime}=y^{\prime} $.

\textbf{Acknowledgments}

Two of the authors (G.S. and G.V.) wish to thank G. Bimonte, B. Dubrovin
and G.Marmo for interesting discussions.

\end{document}